\documentclass{article}
\usepackage{ijcai22}

\usepackage{times}

\usepackage{soul}
\usepackage{url}
\usepackage[hidelinks]{hyperref}
\usepackage[utf8]{inputenc}
\usepackage[small]{caption}
\usepackage{graphicx}
\usepackage{amsmath}
\usepackage{booktabs}
\usepackage{amssymb}
\usepackage{xcolor}

\usepackage{algorithm}
\usepackage{algorithmic}

\newcommand{\bX}{\mathbf{X}}
\newcommand{\bx}{\mathbf{x}}

\newcommand{\bv}{\mathbf{v}}
\newcommand{\cI}{\mathcal{I}}
\newcommand{\MMS}{\mathsf{MMS}}

\newcommand{\PROP}{\mathsf{PROP}}

\newtheorem{theorem}{Theorem}[section]

\newtheorem{definition}[theorem]{Definition}

		\newcommand{\citet}[1]{\citeauthor{#1}~\shortcite{#1}}
		\newcommand{\citep}{\cite}
\title{
Algorithmic Fair Allocation of Indivisible Items: A Survey and New Questions}
\author{
Haris Aziz$^1$\and
Bo Li$^2$\and
Herv\'e Moulin$^{3}$\And
Xiaowei Wu$^4$\\
\affiliations
$^1$UNSW Sydney and Data61 CSIRO, Sydney, Australia\\
$^2$Department of Computing, The Hong Kong Polytechnic University, Hong Kong, China\\
$^3$University of Glasgow, Glasgow, UK and Higher School of Economics, St. Petersburg, Russia\\
$^4$IOTSC, University of Macau, Macau, China\\
\emails
haziz@cse.unsw.edu.au,
comp-bo.li@polyu.edu.hk,
herve.moulin@glasgow.ac.uk,
xiaoweiwu@um.edu.mo
}

\begin{document}

\maketitle

\begin{abstract}
The theory of algorithmic fair allocation is within the center of multi-agent systems and economics in the last decade due to its industrial and social importance. At a high level, the problem is to assign a set of items that are either goods or chores to a set of agents so that every agent is happy with what she obtains. Particularly, in this survey, we focus on indivisible items, for which absolute fairness such as envy-freeness and proportionality cannot be guaranteed. One main theme in the recent research agenda is about designing algorithms that approximately achieve the fairness criteria. We aim at presenting a comprehensive survey of recent progresses through the prism of algorithms, highlighting the ways to relax fairness notions and common techniques to design algorithms, as well as the most interesting questions for future research.
\end{abstract}

\pagestyle{plain}

\section{Introduction}
% Fair allocation has become an important research question in multi-agent systems that integrates computer science and social sciences \cite{books/daglib/0017734}. 
While fair allocation is an age-old problem and the widely known Divide-and-Choose algorithm can trace back to the Bible, modern research on fair allocation is regarded to be initiated by Steinhaus at a meeting of the Econometric Society in Washington D.C. in 1947 \cite{steinhaus1948problem}.
Since then, a large body of works in economics and mathematics have been directed towards understanding the theory of allocating resources among agents in a fair manner \cite{books/daglib/0017734}. 
The recent focus on {\em indivisible} items is motivated, in part, by the applications that inherently entail allocation of items that cannot be fractionally allocated, such as assigning computational resources in a cloud computing environment and courses to teachers in a school. 
In the last decade, computer science offered a fresh and practical angle to the research agenda -- algorithmic fair allocation. 
Besides designing algorithms, computer science has brought many more ideas, such as computational and communication complexity, and informational assumptions, which do not align with the main theme of the current survey. Interested readers can refer to the surveys by \citet{conf/ijcai/Walsh20} and \citet{conf/aaai/Aziz20} for detailed discussion.
%Algorithms developed for finding fair 
Fair allocation algorithms have been implemented in the real world; 
for instance, Course Match is employed for course allocation at the Wharton School in the University of Pennsylvania, and the websites Spliddit (\url{spliddit.org}) and Fair Outcomes (\url{fairoutcomes.com}) provide online access to fair allocation algorithms.

%\footnote{\url{mba-inside.wharton.upenn.edu/course-match/}, \url{spliddit.org}, \url{fairoutcomes.com/}}

Although this survey mainly focuses on indivisible items, the study of fair allocation was classically centered around allocating a divisible resource, which is also known as the {\em cake-cutting problem}  \cite{books/daglib/0017730,books/daglib/0017738}.
Fairness is mostly captured by {\em envy-freeness} and {\em proportionality} in the literature.
An envy-free allocation (which is also proportional) of a divisible cake always exists and can be found in bounded steps \cite{conf/focs/AzizM16}.
Moreover, a competitive equilibrium from equal incomes guarantees envy-freeness and Pareto optimality simultaneously \cite{varian1973equity}. 
A recent line of research extends the study to chores, such as the computation of envy-free allocations \cite{conf/soda/DehghaniFHY18} and competitive equilibria \cite{journals/corr/abs-2107-06649,conf/soda/ChaudhuryGMM21}.
Unlike divisible items, when items are indivisible, absolutely fair allocations rarely exist. 
For example, when allocating a single item to two agents, every allocation is not envy-free or proportional.
Accordingly, an extensively studied subject is to investigate the extent to which these fairness notions or their relaxations can be approximately satisfied.
%by either designing (approximation) algorithms or identifying tricky instances where no algorithm can be better than a certain performance. 

% A traditional fair allocation problem is modeled as allocating a set of resources (abstracted as items) to a set of agents, where every agent's absolute value depends on her own allocated resources, but her relative happiness depends on the comparison of her due share with others. Fairness is mostly captured by envy-freeness and proportionality in the literature. Although an exact envy-free or proportional allocation always exists when the resources are divisible (like funding and land), 

% \haris{While divisible items are not considered, it may be nice to give a few representative refs for work on it such Herve's paper and recent papers by Mehta and others on competitive allocation of divisible chores: SODA 2021, AAMAS 2021 etc. }

There are several surveys highlighting different perspectives of fair allocation theory.
\citet{moulin2018fair} reviewed the theory through the  prism of economics.
%\cite{books/sp/16/LangR16,moulin2018fair,conf/aaai/Aziz20,conf/ijcai/Walsh20,conf/aaai/AleksandrovW20,suksompong2021constraints}.
% A survey through the  prism of economics was provided by \citet{moulin2018fair}.
%\citet{moulin2018fair} reviewed fair allocation from the angle of economics; 
\citet{conf/aaai/AleksandrovW20} and \citet{suksompong2021constraints} respectively focused on the online and constrained settings.
\citet{books/sp/16/LangR16}, \citet{conf/ijcai/Walsh20} and
\citet{conf/aaai/Aziz20} reviewed the problem in the perspective of broad computer science. 
Instead, the angle of the current survey is algorithmic and the focus is particularly on the introduction of 
%various ways to relax fairness requirements and 
common techniques to design (approximation) algorithms.
Moreover, we will discuss more sophisticated settings that were introduced in the last couple of years which 
uncovered new challenges and open problems in the field of fair allocation. 
%which will also be included in the current survey.

% the last couple of years have witnessed the flourish of more sophisticated settings that uncovers new challenges and open problems to fair allocation, which will also be included in the current survey.

\paragraph{Roadmap}
In the remaining of the survey, we define the model of fair allocation in Section \ref{sec:model} and introduce the widely adopted solution concepts in Sections \ref{sec:EF} and \ref{sec:PROPandMMS}.
% For each concept, we review the state-of-the-art results as well as the open problems.
In Section~\ref{sec:algorithms}, we review the commonly used techniques to design fair allocation algorithms.
In Section \ref{sec:extensions}, we introduce the more sophisticated settings that are proposed recently.
Finally, we discuss two more properties, efficiency and truthfulness that are desired to be satisfied together with fairness,  in Section \ref{sec:discussion}.

\section{Model}
\label{sec:model}

In a fair allocation instance, we allocate a set of $m$ indivisible items $M=\{1,\ldots,m\}$ to a group of $n$ agents $N=\{1,\ldots,n\}$. 
% A bundle is a subset of items $X\subseteq M$ and
An allocation is represented by an $n$-partition $\bX = (X_1,\ldots,X_n)$ of $M$, where $X_i\subseteq M$ is the bundle allocated to agent~$i$.
It is required that each item is allocated to exactly one agent, i.e.,
$X_i \cap X_j = \emptyset$ for all $i \neq j$ and $\cup_{i\in N} X_i = M$.
If $\cup_{i\in N} X_i \neq M$, $\bX$ is a {\em partial} allocation. 
We sometimes consider fractional allocations, denoted by $\bx=(x_{ie})_{i\in N, e\in M}$, where $0\le x_{ie} \le 1$ denotes the fraction of item $e$ allocated to agent~$i$, and $\sum_{i\in N}x_{ie} = 1$ for all $e\in M$.
Each agent $i$ has a valuation function $v_i: 2^M \rightarrow \mathbb{R}$ that assigns a value to each bundle of items.
When $v_i(S)\geq 0$ for all $i\in N$ and $S\subseteq M$, the items are \emph{goods}; when $v_i(S) \leq 0$ for all $i$ and $S$, the items are \emph{chores}.
% To better distinguish the two cases, for the allocation of chores we assume that each agent $i \in N$ has a \emph{cost function} $c_i$ that assigns a non-negative \emph{cost} to each bundle of items.
%due to space limit.
For ease of exposition, we mainly discuss the case when the valuations are additive and leave the discussion on more general valuations to Section \ref{sec:extensions}.
That is, for any $i\in N$ and $S \subseteq M$, we have $v_i(S) = \sum_{e\in S} v_i(\{e\})$.
When there is no confusion, we use $v_i(e)$ to denote $v_i(\{e\})$. % for convenience.
Further, for any $S\subseteq M$ and $e\in M$, we use $S+e$ and $S-e$ to denote $S\cup\{e\}$ and $S\setminus\{e\}$, respectively.
Let $\cI = (N,M,\bv)$ be a fair allocation instance where $\bv=(v_1,\ldots,v_n)$.
When all agents agree on the same ordering of all items in values (e.g., $v_{i}(1) \ge \cdots \ge v_{i}(m)$ for all $i$), the instance is called {\em identical ordering} (IDO).
% It was shown by \citet{journals/siamdm/PlautR20} that EFX allocations always exist for arbitrary number of agents with identical ordering (IDO) additive valuations (the valuation functions of all agents agree on the same ordering of items in values)

Before the extensive study of fairness, efficiency was at the centre of the theory of resource allocation. 
The \emph{utilitarian welfare} of an allocation $\bX$ is $\sum_{i\in N} v_i(X_i)$, by maximizing which the total happiness of the agents is maximized. The \emph{egalitarian welfare} is $\min_{i\in N} \{v_i(X_i)\}$, by maximizing which the smallest happiness is maximized.
A compromise between utilitarian and egalitarian welfare is \emph{Nash welfare}, i.e., $\Pi_{i\in N} v_i(X_i)$.
We say an allocation $\bX$ \emph{Pareto dominates} another allocation $\bX'$ if $v_i(X_i) \geq v_i(X'_i)$ for all $i\in N$ and $v_i(X_i) > v_i(X'_i)$ for some $i$.
An allocation is \emph{Pareto optimal} (PO) if it is not Pareto dominated by any other allocation.

Naturally, the fairness of an allocation can be evaluated by its egalitarian welfare, such Santa Claus problem \cite{conf/stoc/BansalS06} and load balancing problem \cite{journals/mp/LenstraST90}. 
However, in practice, since agents may have heterogeneous valuations, the max-min objective is not enough to satisfy all of them.
Thus, various notions were proposed to characterize the fairness of allocations, including envy-freeness (EF) \cite{foley1966resource}, proportionality (PROP) \cite{steinhaus1948problem} and equitability (EQ) \cite{dubins1961cut}.
The relationships among these notions are discussed by \citet{conf/ijcai/AmanatidisBM18},\citet{conf/atal/SunCD21} and \citet{journals/corr/abs-2112-04166}.
Due to space limit, this survey only focuses on two of the most widely studied, namely EF and PROP.
%In this section, we review their definitions, relaxations, algorithmic results and open problems.

%\section{Fairness Solution Concepts}
%\label{sec:definitions}

% \haris{Perhaps say Fairness Concepts as we don't seem to focus on efficiency etc. Also what about leximin and equitability?}

\section{Envy-freeness}
\label{sec:EF}

We first consider envy-freeness, whose study dates back to \citet{foley1966resource} and \citet{tinbergen1930mathematiese}, and its relaxations.
% The notion was introduced for allocating divisible items, and naturally extends to indivisible items.
%\haris{There seem to be older references to envy that are not by Foley: https://ejpe.org/journal/article/view/610 . Apparently, Tinbergen introduced ‘no-envy’ as a fairness criterion in his article “Mathematiese Psychologie” published in 1930 in the Dutch journal Mens en Maatschappij and translated as “Mathematical Psychology” in 2021 in the Erasmus Journal for Philosophy and Economics. }

\begin{definition}[EF]
	For the allocation of items (goods or chores), an allocation $\bX$ is envy-free (EF) if for any two agents $i,j\in N$, we have $v_i(X_i) \geq v_i(X_j)$.
	%(resp. $c_i(X_i) \leq c_i(X_j)$).
\end{definition}

% In contrast to the case of allocating divisible items, for which EF allocations always exist~\cite{brams1995envy}, when the items to be allocated are indivisible, EF allocations do not always exist.

%As an EF allocation may not exist, t
The problem of checking whether a given instance admits an EF allocation is NP-complete even for $\{0,1\}$- or $\{0,-1\}$-valued instances \cite{DBLP:journals/ai/AzizGMW15,conf/approx/BhaskarSV21}.
Moreover, the example of allocating a single item between two agents defies any bounded multiplicative approximation of EF, and thus researchers turn their attention to additive approximations.
Two of the most popular ones are envy-free up to one item (EF1) and envy-free up to any item (EFX).

\paragraph{EF1}
The notion of EF1 was first studied for the allocation of goods by \citet{conf/sigecom/LiptonMMS04}, which allows an agent to envy another agent but requires that the envy can be eliminated by removing an item from the envied agent's bundle.
This notion naturally extends to chores by removing an item from the envious agent's bundle.
For both goods and chores, EF1 allocations always exist and can be efficiently computed by the Round Robin algorithm; see Section \ref{sec:algorithms}.
% The notion naturally extends to the allocation of chores.
% Moreover, it is straightforward to show that for the allocation of chores the Round Robin algorithm computes an EF1 allocation in polynomial time.

\begin{definition}[$\alpha$-EF1]
For any $\alpha \geq 0$, an allocation $\bX$ is $\alpha$-approximate envy-free up to one item ($\alpha$-EF1) if for any $i,j \in N$, there exists $e \in X_i\cup X_j$ such that $v_i(X_i - e)\geq \alpha\cdot v_i(X_j - e)$.
When $\alpha = 1$, the allocation is EF1.
\end{definition}

\paragraph{EFX}
Same as EF1, the EFX relaxation was proposed for the allocation of goods, by \citet{journals/teco/CaragiannisKMPS19}.
Informally speaking, the notion of EFX strengthens the fairness by requiring that the envy between two agents can be eliminated by removing \emph{any} item owned by these two agents.

\begin{definition}[$\alpha$-EFX]
	For any $\alpha \ge 0$, an allocation $\bX$ for goods (resp. chores) is $\alpha$-approximate envy-free up to any item ($\alpha$-EFX) if for any $i,j \in N$ and any $e \in X_j$ (resp. $e\in X_i$), $v_i(X_i)\geq \alpha\cdot v_i(X_j - e)$ (resp. $v_i(X_i - e)\geq \alpha\cdot v_i(X_j)$).
	When $\alpha = 1$, the allocation is EFX.
\end{definition}

Unlike the case of EF1 allocations, 
%which always exist and can be efficiently computed, 
the existence of EFX allocations remains unknown.
%even for very special cases.
% The existence and computation of (approximately) EFX allocation of goods have attracted considerable attention in the past few years.
For the case of goods, it was shown by \citet{journals/siamdm/PlautR20} that EFX allocations exist in some special cases:
(1) identical (combinatorial) valuations, (2) IDO additive valuations, and (3) $n=2$. 
% arbitrary number of agents with identical ordering (IDO) additive valuations (the valuation functions of all agents agree on the same ordering of items in values); and the case of two agents with general (non-additive) valuation functions.
\citet{conf/sigecom/ChaudhuryGM20} and \citet{journals/tcs/AmanatidisBFHV21} further extended the existence of EFX allocations to the cases when (4) $n=3$, and (5) bi-valued valuations. 
% \citet{journals/tcs/AmanatidisBFHV21} consider the bi-valued instances, for which there exists $\epsilon\in [0,1)$ such that $v_i(e) \in \{ 1,\epsilon \}$ for all $i\in N$ and $e\in M$, and propose a polynomial time algorithm for computing EFX allocations.
%\paragraph{EFX Partial Allocations for Goods.}
While the existence of EFX allocations remains unknown for the general cases, there are fruitful results regarding EFX partial allocations (where unallocated items are assumed to be donated to a \emph{charity}) and approximation of EFX allocations.
% Recall that a partial allocation $\bX$ is a partition of a proper subset $M'\subseteq M$ into $n$ bundles, where $M' = \bigcup_{i\in N} X_i$ is the set of allocated items and $P = M\setminus M'$ is the set of \emph{unallocated} items.
% The set of items in $P$ are often referred to as items donated to the \emph{charity}.
Since allocating nothing to the agents is trivially EFX, researchers are interested in finding EFX partial allocations with high efficiency.
%, or those with few unallocated items. 
\citet{conf/ec/CaragiannisGH19} showed that there exists an EFX partial allocation achieving half of the maximum Nash welfare.
\citet{journals/siamcomp/ChaudhuryKMS21} proposed a pseudo-polynomial time algorithm that computes an EFX partial allocation with at most $n-1$ unallocated items under which no agent envies the charity.
This result was improved by \citet{journals/corr/abs-2102-10654}, who showed that there is an EFX allocation with at most a single unallocated item for $n=4$, and $n-2$ unallocated items for $n\geq 5$.
% 
%Besides \cite{conf/sigecom/ChaudhuryGMMM21},
%\paragraph{Approximately EFX Allocations for Goods.}
There are also results that aim at computing approximately EFX allocations.
\citet{journals/siamdm/PlautR20} showed that every instance (even with subadditive valuations) admits a $0.5$-EFX allocation.
%and \citet{conf/ijcai/Chan00W19} proposed a polynomial time algorithm to compute one.
The approximation ratio was improved to $0.618$ under additive valuations by a polynomial time algorithm proposed by \citet{journals/tcs/AmanatidisMN20}.
\citet{conf/sigecom/ChaudhuryGMMM21} proposed a polynomial time algorithm that computes a $(1-\epsilon)$-EFX allocation with $o(n)$ unallocated items and high Nash welfare.
% 
%\paragraph{(Approximately) EFX Allocations for Chores.}
% 
In contrast to the case of goods, the chores counterpart is much less well studied.
EFX allocations are known to exist only for a few special cases, e.g., IDO instances~\cite{journals/corr/abs-2103-11849} and leveled preference instances~\cite{journals/corr/abs-2109-08671}.
For general instances, only $O(n^2)$ approximation of EFX is known to exist \cite{journals/corr/abs-2109-07313}.
The existence of EFX allocations remains unknown even for $n=3$ agents or bi-valued instance.

\section{Proportionality and Maximin Share}
\label{sec:PROPandMMS}

Proportionality (PROP) was
%another widely adopted fairness criterion, 
proposed by \citet{steinhaus1948problem},
%in the context of a divisible resource.
which is the most widely studied threshold-based solution concept.
PROP is weaker than EF under additive valuations.
% The notion is formally defined as follows. 

\begin{definition}[PROP]
	An allocation $\bX$ is proportional (PROP) if for every agent $i\in N$, we have $v_i(X_i) \geq \PROP_i$, where $\PROP_i = (1/n) \cdot v_i(M)$.
\end{definition}

For divisible goods and normalized valuations, the items can be allocated such that every agent has value at least $1/n$, which is not true for indivisible items.
%, e.g., consider allocating a single item to two agents.
\citet{hill1987partitioning} studied the worst
case guarantee that an agent can have as a function of $n$ and $\max_{i\in N,e\in M} \{v_{i}(e)\}$. 
With two agents, the chores version is equivalent to the goods one;
but with three or more agents, the equivalence is far from clear, and may not hold. 
One drawback of this guarantee is that the value of the function decreases quickly and goes to 0 as $\max_{i,e} \{v_{i}(e)\}$ becomes large.

\paragraph{Maximin Share Fairness}

Besides the worst case guarantee studied by \citet{hill1987partitioning}, one popular relaxation of PROP is the maximin share fairness, motivated by the following imaginary experiment.
% A PROP allocation requires that every agent $i$'s value to be no smaller than $1/n\cdot v_i(M)$, and at the same time the agents also know that PROP cannot be guaranteed even if they are the mediator. 
%Consider the following imaginary experiment where 
If agent $i$ is the mediator and divides all items into $n$ bundles, the best way to approximate PROP for~$i$ is to maximize the smallest bundle according to her own valuation.
Formally, define the maximin share (MMS) of $i$ as 
\[
\MMS_i(M,n) = \max_{\bX \in \Pi_n(M)} \min_j \{v_i(X_j)\},
\]
where $\Pi_n(M)$ denotes the set of all $n$-partitions of $M$.
When $M$ and $n$ are clear from the context, we write $\MMS_i$ for short.
Note that $\MMS_i \le \PROP_i$, and the computation of $\MMS_i$ is NP-complete.

\begin{definition}[$\alpha$-MMS]
	For any $\alpha \geq 0$, an allocation $\bX$ is $\alpha$-approximate maximin share fair ($\alpha$-MMS) if for any $i \in N$, $v_i(X_i)\geq \alpha\cdot \MMS_i$.
	%and $v_i(X_i)\leq \alpha\cdot \MMS_i$ for chores.
	When $\alpha = 1$, the allocation is MMS.
\end{definition}
% Accordingly, the requirement of fairness is relaxed to that every agent $i$'s value is no less than $\MMS_i$, which is called {\em  maximin share} (MMS) fair.
% With additive valuations, an EF allocation is always PROP, and thus is also MMS fair.
Note that the approximation ratio $\alpha \leq 1$ for goods and $\alpha \geq 1$ for chores.
The definition of MMS fairness was first introduced by \citet{journals/bqgt/Budish10}, based on the concept of \cite{moulin1990uniform}.
Unfortunately, it is shown that there exist instances for which no allocation can ensure $\MMS_i$ value for every agent for the case of goods~\cite{journals/jacm/KurokawaPW18} and chores~\cite{conf/aaai/AzizRSW17}.
% 
% Constant upper bounds are provided in \cite{journals/corr/abs-2104-04977}. 
% Since exact MMS fairness cannot be guaranteed for both goods and chores, people focus on their approximations.
The best known approximation results are $(3/4+1/(12n))$-MMS for goods \cite{journals/ai/GargT21} and $11/9$-MMS for chores \cite{conf/sigecom/HuangL21}.
The best known negative results are that $\alpha \le 39/40$ for goods and for $\alpha \ge 44/43$ for chores \cite{journals/corr/abs-2104-04977}.

\paragraph{More Solution Concepts}

% In the following we introduce more solution concepts that are closely related to MMS and PROP.
Motivated by the definition of MMS, \citet{journals/teco/CaragiannisKMPS19} proposed {\em pairwise MMS} (PMMS) and \citet{conf/aaai/BarmanBMN18} proposed {\em groupwise MMS} (GMMS).
% \begin{definition}[$\alpha$-PMMS and $\alpha$-GMMS]
%     For all $\alpha \geq 0$, an allocation $\bX$ is $\alpha$-approximate pairwise MMS fair ($\alpha$-PMMS) if $\forall i,j \in N, v_i(X_i)\geq \alpha\cdot \MMS_i(X_i\cup X_j,2)$.
%     % \[
%     % v_i(X_i)\geq \alpha\cdot \max_{(A_i,A_j)\in \Pi_2(X_i\cup X_j)} \min\{v_i(A_i),v_i(A_j)\}.
%     % \]
%     It is $\alpha$-approximate groupwise maximin share fair ($\alpha$-GMMS) if $\forall i \in N$, $v_i(X_i)\geq \alpha\cdot \max_{S \subseteq N: S \ni i} \{\MMS_i(\cup_{j\in S} X_j, |S|)\}$.
%     % \[
%     % v_i(X_i)\geq \alpha\cdot \max_{S \subseteq N, S \ni i, \bA\in \Pi_{|S|}(\cup_{j\in S} X_j)} \min_{j\in S}\{v_i(A_j)\}.
%     % \]
% 	When $\alpha = 1$, the allocation $\bX$ is PMMS or GMMS.
% \end{definition}
Informally, PMMS is similar to MMS, but instead requires that the allocation is MMS for the instance induced by any two agents.
% of partitioning all items into $n$ bundles, an agent partitions the union of items allocated to her and another agent into two bundles, and receives the less valued one.
GMMS generalizes both MMS and PMMS and requires that the allocation is MMS for the instance induced by any subset of agents.
% the max-min value on the union of any subset of agents.
We refer the readers to, e.g., \cite{journals/teco/CaragiannisKMPS19,conf/aaai/BarmanBMN18,journals/tcs/AmanatidisMN20}, for more detailed discussions.

% \cite{conf/aaai/BarmanBMN18}.
% The PMMS fairness notion is similar to MMS in that it requires the allocation to be MMS for every instance induced by the allocation $\bX$ and any two agents.
% % she partitions the combined bundle of herself and another player into two bundles, and receives the one she values less. 
% It is noted in \cite{journals/teco/CaragiannisKMPS19} that neither the PMMS guarantee nor the MMS guarantee implies the other, but a PMMS allocation is always $0.5$-MMS.
% % 
% Beyond the MMS related notions, we can also adapt the ``up to one/any'' relaxation to proportionality. 

% \begin{definition}[PROP1 and PROPX]
% 	An allocation $\bX$ is called proportional up to one item (PROP1) if for all $i\in N$, we have
% 	$v_i(X_i\cup\{e\}) \geq \PROP_i$ for some good $e\notin X_i$ or $v_i(X_i\setminus\{e\}) \geq \PROP_i$ for some chore $e\in X_i$.
% 	The allocation is called proportional up to any item (PROPX) if for all $i\in N$, we have
% 	$v_i(X_i\cup\{e\}) \geq \PROP_i$ for any good $e\notin X_i$ or $v_i(X_i\setminus\{e\}) \geq \PROP_i$ for any chore $e\in X_i$.
% \end{definition}

Finally, similar to EF1 and EFX, we can relax PROP to PROP1 and PROPX. 
It is known that a PROP1 allocation always exists and can be found in polynomial time when the items are goods \cite{conf/sigecom/ConitzerF017,conf/aaai/BarmanK19}, chores \cite{journals/corr/abs-1907-01766} or mixture of goods and chores \cite{journals/orl/AzizMS20}.
% Moreover, all these results compute allocations that are both PROP1 and PO.
Regarding PROPX, when items are goods, PROPX allocations may not exist \cite{moulin2018fair,journals/orl/AzizMS20}.
However, when items are chores, PROPX allocations exist and can be found efficiently \cite{moulin2018fair,journals/corr/abs-2103-11849}. 

\section{Algorithms and Common Techniques}
\label{sec:algorithms}

In this section, we introduce the techniques to design fair allocation algorithms. 
Unfortunately, due to space limit, we were not able to include all of them.
Instead, we choose some of the most commonly used and powerful ones that are also the basis of more complicated algorithms.

\paragraph{Divide-and-Choose}
Divide-and-Choose is one of the most classic allocation algorithms.
%which traces back to the Bible.
The algorithm is very useful and intuitive when there are only two agents.
The idea is to let the first agent partition the items into two bundles and the other agent choose her preferred bundle.
The remaining bundle is allocated to the first agent, and thus her best strategy is to maximize the value of the smaller bundle, i.e.,
\begin{equation*}
     (X_1,X_2) \in \arg \max\limits_{(S_1,S_2) \in \Pi_2(M)} \min \{v_1(S_1),v_1(S_2)\}.
\end{equation*}
\citet{journals/siamdm/PlautR20} proved that $(X_1,X_2)$ is always MMS and EFX to the first agent if we break tie by maximizing the size of the smaller bundle in $(X_1,X_2)$, i.e., leximin++ allocation. %as proved by \citet{journals/siamdm/PlautR20}.
Actually, this result holds for any number of agents which implies the existence of EFX allocations for the case of identical valuations. 
Since the second agent obtains her preferred bundle in $(X_1,X_2)$, the allocation is EF to her.
Therefore, with two agents, Divide-and-Choose algorithm returns an allocation that is MMS and EFX.

% \begin{algorithm}[H]
% 	\caption{\hspace{-2pt}~ $\DC(\{1,2\},M,v)$}
% 	\label{alg:unrelated}
% 	\begin{algorithmic}[1]
% % 	\REQUIRE A randomly permuted stream of $n$ jobs, $m$ machines.
% % 	\ENSURE An assignment $\mathbf s\in\{0,\ldots,m\}^J$
	
% 	\STATE $(M_1,M_2) \in \arg \max\limits_{(S_1,S_2) \in \Pi_2(M)} \min \{v_1(S_1),v_1(S_2)\}$. \label{step:dc:1}
% 	\STATE Let $X_2 \in \arg \max \{v_2(M_1),v_2(M_2)\}$, $X_1 = M\setminus X_2$.
% % 	\FOR{each integer $l\geq 0$, }
% % 	\STATE 
% % 	\ENDFOR
% 	\end{algorithmic}
% \end{algorithm}

\paragraph{Adjusted-Winner}
Adjusted-Winner is another widely used algorithm for the two-agent case \cite{books/daglib/0017730}. 
The idea is to sort the items according to the ratios between the utilities that they yield for the two agents, i.e., 
\begin{equation*}
     \frac{v_{1}(1)}{v_{2}(1)} \ge \frac{v_{1}(2)}{v_{2}(2)} \ge \cdots \ge \frac{v_{1}(m)}{v_{2}(m)},
\end{equation*} 
and let agent 1 choose a minimal set of consecutive items for which she is EF1 starting from left (the remainings are given to agent 2).
The advantage is that it ensures high social welfare between two agents \cite{journals/mst/BeiLMS21}.
This allocation is EF1 but not necessarily MMS or EFX.

% \begin{itemize}
% \item EFX for general monotone identical cost functions~\cite{journals/corr/abs-2006-04428} 
% \end{itemize}

\paragraph{Sequential Allocation and Round-Robin}
% \label{sec:alg:sa+rr}

A general class of algorithms that is also suitable for a distributed implementation is that of sequential-picking allocation \cite{books/daglib/0017729}, which was formally studied in a general and systematic way by \citet{conf/ijcai/BouveretL11}.
Under these methods, agents have a sequence of turns to pick their most preferred item that is still available. 
A popular sequence protocol is the {\em Round-Robin}, where the picking sequence repeats the pattern $1,\ldots, n$. 
The Round-Robin algorithm produces allocations that are EF1 (but not necessarily EFX) for both goods and chores\footnote{When items are only goods or only chores, there is a larger class of protocols ensuring EF1. This class of protocols uses a \textit{recursively balanced} sequence in which at any point, the difference between the number of turns of any of two agents is at most 1.}, but not for mixture of them. For this, \citet{journals/orl/AzizMS20} proposed the double Round-Robin method that computes EF1 allocations for mixture of goods and chores. 
\citet{conf/ijcai/AmanatidisBM16} and \citet{journals/corr/abs-2012-13884} designed more involved picking sequences to approximate MMS fairness, for goods and chores respectively.

% Though Round-Robin does not have a good guarantee for MMS with goods, it ensures $2$-MMS for chores \cite{conf/aaai/AzizRSW17}.
% Later, \citet{journals/corr/abs-2012-13884} improved this approximation ratio to $5/3$ via a more involved picking sequence.

% \bo{Shall we say anything about incentives like the following? Although the focus of the survey is on designing fair algorithms, incentives in sequential-picking algorithms has attracted considerable attention. For example, ...}

% \haris{I suggest `sequential allocation and have it separate from envy-cycles elimination'. We know that round robin sequential allocation satisfies EF1 for goods and also for chores. However, it doesn't for goods and chores. For this, \cite{journals/orl/AzizMS20} proposed for the double round robin method. When talking about sequential allocation, you can also discuss interesting sequences that give good MMS approximation guarantees. Some sample text: is below}

% \harisnew{A general class of algorithms that is also suitable for a distributed implementation is that of sequential allocation (see, e.g. Brams XXX,  Bouveret and Lang (2011), Brams ). Under these methods, agents have a sequence of turns and in their turn they pick the most preferred that is still available. A popular sequence protocol is to have agents come in a round robin manner. When items are only goods or only chores, a larger class of protocols ensures EF1. This class of protocols uses a \textit{recursively balanced} sequence in which at any point, the difference between the number of turns of any of two agents is at most 1.}

\paragraph{Envy-cycle Elimination}

The Envy-cycle Elimination algorithm is inherently a greedy algorithm that in each round a new item is assigned to the agent who is in a disadvantage for goods or advantage for chores \cite{conf/sigecom/LiptonMMS04}.
The main technique of the algorithm is to ensure the existence of an agent that is not envied (for goods) or not envious (for chores) by trading items among agents.
We use goods as an illustration. 
The algorithm is based on an {\em envy graph}, where the nodes correspond to agents and there is an edge from agent $i$ to agent $j$ if $i$ is envious of $j$'s bundle. 
The algorithm works by assigning, at each step, an unassigned item to an agent who is not envied by any other agents, 
i.e., a node with in-degree $0$ in the envy graph. 
%i.e., node correspond to agent $i$ with in degree $0$. 
If no such agent exists, the graph must contain a directed cycle.
Then the cycle can be resolved by exchanging the bundles of items along the cycle, i.e., an agent in the cycle gets the bundle of the agent she points to.
The algorithm terminates when all items are allocated and outputs an EF1 allocation for arbitrary monotone combinatorial valuation functions \cite{conf/sigecom/LiptonMMS04}. 

The algorithm and its adaptations are very widely studied, combining with which stronger fairness notions can also be satisfied.
% 
% For example, consider the allocation of goods.
For example, the algorithm itself ensures EFX \cite{journals/siamdm/PlautR20} and $2/3$-MMS \cite{journals/teco/BarmanK20} for IDO instances.
% Under sub-additive valuations, combining the algorithm with maximum matching gives a polynomial time algorithm that computes $0.5$-EFX allocations~\cite{conf/ijcai/Chan00W19}. 
With more involved preprocessing procedures, it can ensure 0.618-EFX, 0.553-GMMS,  0.667-PMMS and EF1 simultaneously \cite{journals/tcs/AmanatidisMN20}.
For chores, it is shown in \cite{journals/teco/BarmanK20} that the returned allocation is $4/3$-MMS. % for IDO instances.
However, noted by \citet{conf/approx/BhaskarSV21}, this allocation may not be EF1 if the cycle is resolved arbitrarily.
Instead, they used the top-trading technique (in which each agent only points to the agent she envies the most) to preserve EF1. 
Later, \citet{journals/corr/abs-2103-11849} further showed that with this technique, the returned allocation is PROPX. % for IDO instances. 
We can also observe the shadow of the algorithm in more complicated techniques, such as the (group) champion graphs and rainbow cycle number \cite{conf/sigecom/ChaudhuryGM20,conf/sigecom/ChaudhuryGMMM21} which enable stronger existence and approximation results of EFX.

% \cite{journals/teco/CaragiannisKMPS19}, 

% Techniques: 
% \begin{itemize}
% 	\item Envy-graph
% 	\item Combination of Envy-cycle Elimination and other techniques (matching, rainbow cycle number \cite{conf/sigecom/ChaudhuryGMMM21})
% 	%\item $(1-\epsilon)$-EFX with $o(n)$ unallocated items~\cite{conf/sigecom/ChaudhuryGMMM21}

% 	\item $0.618$-EFX~\cite{journals/tcs/AmanatidisMN20}
% 	\item $4/3$-MMS~\cite{conf/sigecom/BarmanM17,journals/teco/BarmanK20} for chores
% 	\item Existence of Weighted EF1 and PO~\cite{conf/atal/ChakrabortyISZ20}
% 	\item EFX for IDO, weighted PROPX computation, ordinal PROPX and POF \cite{journals/corr/abs-2103-11849}

% \end{itemize}

\paragraph{Bag-filling Algorithms}
% \label{sec:alg:bag-filling}

Bag-filling Algorithms are particularly helpful for the % requirement of
threshold-based fairness like MMS.
%or generally threshold-based fairness notions.
The idea is to maintain a bag and keep adding items to it until some agent thinks the bag is good enough (for goods) or about to be too bad to all agents (for chores).
Then the bag is taken away by some satisfied agent and the algorithm repeats the procedure with the remaining items.  
The difficulty is to select a proper threshold for the bag so that the approximation for the agent who takes away a bag is good and there remains sufficiently many (or few) items for the remaining agents.
% As a concrete example, consider the case of goods when every agent has value at most $0.5\cdot\MMS_i$ on every item. 
% To get a $0.5$-MMS allocation, we stop filling the bag once some agent~$i$ values the bag for no smaller than $0.5\cdot\MMS_i$. 
% Since the other agents are not satisfied before adding the last item, the bag has value less than $\MMS_j$ to every agent $j$.
% This ensures that the remaining agent still have enough items to be satisfied. 
% If there is an agent $i$ who values a single item for greater than $0.5\cdot \MMS_i$, then by allocating the item to $i$, she is satisfied. It can be shown that all other agents are only happy, e.g., the MMS share of every other agent in the induced instance is no smaller~\cite{journals/jacm/KurokawaPW18}.
% 
With a more careful design and analysis, % and analysis of the procedure,  
the approximation ratio can be improved to $2/3$~\cite{conf/soda/GargMT19} and further better than $3/4$~\cite{journals/ai/GargT21} for goods,
% These algorithms focus on IDO instances and use Property~\ref{property:IDO} to generalize to non-IDO instances.
% 
For chores the approximation ratio can be improved to $11/9$ \cite{conf/sigecom/HuangL21}.
There are several nice properties regarding MMS fairness \cite{journals/talg/AmanatidisMNS17,conf/soda/GargMT19}, e.g., the scale invariant and the reduction to IDO instances.
Interestingly, the second property shows that any algorithm for approximating MMS allocations for IDO instances applies to general instances with the same approximation ratio preserved.
Making use of these properties the design of algorithms can be significantly simplified. 

\paragraph{Rounding Fractional Solutions}

% \haris{Perhaps we can combine fractional and rounding and competitive and then remark that some algorithms only round. }
%\subsection{Competitive Equilibrium}
% \haris{Some rough notes: please edit as you please}
% \harisnew{
Although competitive equilibria may not exist for indivisible items, 
% is known to exist for divisible items~\cite{RePEc:wly:emetrp:v:85:y:2017:i:6:p:1847-1871} but may not exist for indivisible items. 
we can first compute a market equilibrium by assuming the items being divisible and then carefully round the fractional allocation to an integral one~\cite{conf/aaai/BarmanK19,journals/corr/abs-1907-01766,conf/sagt/GargHMS21}.
% A general approach to deriving desirable or approximately competitive equilibrium involves computing competitive fractional allocations and carefully rounding them to integral allocations~\cite{conf/aaai/BarmanK19,journals/corr/abs-1907-01766,conf/sagt/GargHMS21}.
This approach is especially helpful when efficiency is desired along with fairness, e.g., for
the computation of 
EF1+PO or PROP1+PO allocations for goods \cite{conf/sigecom/BarmanKV18,conf/aaai/BarmanK19},
EF1+PO allocations for bi-valued chores \cite{journals/corr/abs-2110-09601,journals/corr/abs-2110-11285},
PROP1+PO \cite{journals/orl/AzizMS20} and
approximately MMS+PO allocations for mixture of goods and chores \cite{conf/sigecom/KulkarniMT21}.
% }

% FPTAS for computing CEEI for divisible chores~\cite{journals/corr/abs-2107-06649} 

\paragraph{Eating Algorithms}

The Probabilistic-Serial (PS) algorithm of~\cite{journals/jet/BogomolnaiaM01} is a randomized algorithm for allocating indivisible items in an ex-ante EF manner. 
Agents eat their most preferred items at a uniform rate and move on to the next item when the previous one is consumed. The probability share of an agent for an item is the fraction of the item eaten by the agent. 
In recent works \cite{conf/sigecom/Freeman0V20,conf/wine/Aziz20}, researchers have sought allocation algorithms that simultaneously satisfy ex-ante EF and ex-post EF1 for the allocation of indivisible items that are goods or chores. In particular, the PS-lottery method was proposed that provides an explicit lottery over a set of EF1 allocations.
\citet{journals/corr/abs-2008-08991} presented an eating algorithm that is suitable for any type of feasibility constraint and allocation problem with ordinal preferences.

\section{More Sophisticated Settings}
\label{sec:extensions}

The past several years have witnessed the emergence of more sophisticated settings that brought new challenges to the design of fair allocation algorithms, including mixture of goods and chores, weighted agents, partial information and general valuations.
In the following, we review their models, as well as the corresponding results and open problems. %, for some of them.

\paragraph{Mixture of Goods and Chores}
The general case when items are mixture of goods and chores has recently been studied in \cite{BogomolnaiaMSY17,journals/scw/BogomolnaiaMSY19a,journals/orl/AzizMS20,journals/aamas/AzizCIW22}.
This model is particularly interesting because it includes the typical setting when the valuations are not monotone. 
\citet{journals/aamas/AzizCIW22} proved that a double Round Robin algorithm is able to compute an EF1 allocation for any number of agents, and a generalized adjusted winner algorithm can find an EF1 and PO allocation for two agents. 
A natural open question is whether PO and EF1 allocations exist for arbitrary number of agents.
Recently, \citet{journals/orl/AzizMS20} and \citet{conf/sigecom/KulkarniMT21} designed algorithms that compute PROP1+PO or approximately MMS+PO allocations, respectively.
%for mixture of indivisible goods and chores.
More generally, it is an intriguing future research direction to study the fair allocation problem under arbitrary non-monotonic valuations. 

% For indivisible mixtures, \cite{journals/orl/AzizMS20} designed an algorithm to compute a PROP1 and PO allocation.

% Most of the existing works are focused on the competitive equilibrium when items are divisible \cite{RePEc:wly:emetrp:v:85:y:2017:i:6:p:1847-1871,journals/scw/BogomolnaiaMSY19a,conf/soda/ChaudhuryGMM21}. 

\paragraph{Asymmetric Agents}
For most of the aforementioned research works, the agents are assumed to be symmetric in the sense of taking the same share in the system. 
Motivated by the real-world scenarios where people at leadership positions take more responsibilities, some recent works took the asymmetry into consideration and studied the fair treatment of non-equals. %weighted fairness. % who may have different shares. 
The definitions of envy-freeness and maximin share fairness have been adapted to the weighted settings by \citet{journals/jair/FarhadiGHLPSSY19}, \citet{conf/ijcai/0001C019} and \citet{conf/atal/ChakrabortyISZ20}.
Regarding goods, it is shown by \citet{journals/jair/FarhadiGHLPSSY19} that the best approximation ratio for weighted MMS is $\Theta(n)$ and by \citet{conf/atal/ChakrabortyISZ20} that weighted EF1 allocations exist. 
Regarding chores, although weighted MMS was studied in \cite{conf/ijcai/0001C019}, the best approximation ratio and the existence of weighted EF1 allocations are still unknown. 
Novel fairness notions, such as AnyPrice share and $l$-out-of-$d$ maximin share, were proposed and studied in \cite{conf/sigecom/BabaioffEF21,journals/mor/BabaioffNT21} 
which highlight different perspectives of the weighted setting.

\paragraph{With Monetary Transfers}
Since fairness notion like envy-freeness cannot be satisfied exactly, there are works studying how to use payments or subsidies to compensate agents and achieve fairness accordingly \cite{conf/sagt/HalpernS19,conf/sigecom/BrustleDNSV20}. %(i.e., a divisible homogeneous good) 
The problem has been extensively considered in the economics literature under the context of rent division problem \cite{edward1999rental}.
\citet{conf/sagt/HalpernS19} aimed at bounding the amount of external subsidies when the marginal value of each item is at most one for every agent, and \citet{conf/sigecom/BrustleDNSV20} proved that at most one unit of subsidies per agent is sufficient to guarantee the existence of an envy-free allocation \cite{conf/sigecom/BrustleDNSV20}.
\citet{conf/wine/CaragiannisI21} studied the optimization problem of computing allocations that are envy-freeable using minimum amount of subsidies, and designed a fully polynomial time approximation scheme for instances with a constant number of agents.
A more general problem is the fair allocation of mixture of divisible and indivisible items, where the divisible item can be viewed as heterogeneous subsidies \cite{journals/ai/BeiLLLL21,journals/aamas/BeiLLW21}.

\paragraph{Partial Information}
Researchers also care about fair allocation with partial information, and particularly the ordinal preference setting, 
where the algorithm only knows each agent's ranking over all items without the cardinal values.
For goods, the best possible approximation ratio of MMS allocations using only ordinal preferences is $\Omega(\log n)$ by \citet{conf/ijcai/AmanatidisBM16} and \citet{conf/ijcai/0002021};
for chores, constant upper and lower bounds are proved by \citet{journals/corr/abs-2012-13884}. %design algorithms to return approximate MMS allocations for both goods and chores. 
Recently, \citet{DBLP:conf/aaai/HosseiniSVX21} proposed the ordinal MMS fairness, which is more robust to cardinal values. % a more robust approximation notion.
Another interesting question is to investigate the query complexity of unknown valuations.
In this model the algorithms can access the valuations by making queries to an oracle.
\citet{journals/siamdm/OhPS21} proved that $\Theta(\log m)$ queries suffices to define an algorithm that returns EF1 allocations. 
In general, it is an important research direction to explore how much knowledge is sufficient to design a fair allocation algorithm. 

\paragraph{General Valuations}
Besides additive valuations, we may have more complex and combinatorial preferences that involve substitutabilities and complementarities in the items, including submodular, XOS, and subadditive valuations. 
% Then the value for a bundle may be higher or lower than the sum of each individual item’s value.
Formal definitions and discussions of these valuations can be found in \cite{conf/sigecom/Nisan00}.
% (complementarity) or less (substitutability).
% 
Some of the results we have discussed in previous sections also apply to general valuations.
For example, the envy-cycle elimination algorithm returns an EF1 allocation for monotone combinatorial valuations \cite{conf/sigecom/LiptonMMS04}. 
\citet{journals/siamdm/PlautR20} proved the existence of $0.5$-EFX allocations for subadditive valuations.
Regarding MMS, \citet{journals/teco/BarmanK20} and \citet{conf/sigecom/GhodsiHSSY18} designed polynomial time algorithms to compute constant-approximate allocations for submodular and XOS valuations, and $O(\log n)$-approximate for subadditive valuations. 
\citet{conf/aaai/ChaudhuryGM21} further designed algorithms to compute allocations that are approximately EFX and simultaneously achieve $O(n)$-approximation to the maximum Nash welfare for subadditive valuations.

% \paragraph{Fair Division with Constraints}
% Existing works have considered the problem when the allocation needs to satisfy combinatorial constraints.
% One of the most widely studied constraints is {\em connectivity} where the items are assumed to be connected by a graph and the allocation to one agent should be a connected component \cite{conf/ijcai/BouveretCEIP17,conf/aaai/BeiILS21}. 
% Some other constraints include matroid~\cite{conf/aaai/BiswasB19,conf/aaai/DrorFS21}, cardinality~\cite{conf/ijcai/BiswasB18,journals/corr/abs-2106-07300}, budget-feasible \cite{conf/ijcai/00010G21}, conflicting~\cite{journals/aamas/HummelH22}, interval scheduling~\cite{li2021fair} and more.
% We refer the readers to a recent survey by \citet{suksompong2021constraints} for a more detailed discussion of the state-of-the-art results and future directions for the constrained setting.

\smallskip

%\paragraph{Others}
Besides the above settings, there are more in the literature, such as
{\em constrained resources}
\cite{suksompong2021constraints},
%\cite{conf/ijcai/BouveretCEIP17,conf/aaai/DrorFS21},
{\em public resources} \cite{conf/sigecom/ConitzerF017,conf/sigecom/FainM018}, 
{\em group fairness} \cite{journals/mss/Suksompong18,conf/aaai/ConitzerF0V19}, 
and {\em dynamic settings}
\cite{conf/aaai/AleksandrovW20}.
%\cite{conf/sigecom/BenadeKPP18,conf/sigecom/ZengP20}. 
Due to space limit, we were not able to discuss all of them in details.
% For constrained and dynamic settings, interested readers can also refer to the surveys by \citet{suksompong2021constraints} and \citet{conf/aaai/AleksandrovW20}.

\section{Beyond Fairness: Efficiency and Incentives}
\label{sec:discussion}

% Although the problem of fair allocation of indivisible items has become unprecedentedly flourishing, there are still many open problems. 
% As we have mentioned previously, the existence of EFX allocations (for either goods or chores) is still unknown and remains one of the biggest open problem in this field.
% It is also interesting to investigate the best possible approximation of MMS allocations.
% Beyond the above open problems, there are many fascinating future directions and we present some of them as follows.
% \bo{Can we move this to the end of each model?}

Beyond achieving fairness alone, more and more attention is paid to investigating the extent to which we can design algorithms to compute allocations that are fair and simultaneously satisfy other properties, such as efficiency and truthfulness. 

% Besides the open problems that we have discussed while reviewing the corresponding models and unknown results, there are many research directions that are worth effort. 
% Our review has been standing at the perspective of computer science, and in this section, we revisit the fair allocation problem jointly considering another two outstanding perspectives.

\subsection{Computing Fair and Efficient Allocations}
%Before the extensive study of fairness, efficiency was at the centre of the theory of resource allocation. 
% Some common efficiency notions include Nash welfare ($\prod_i v_i(X_i)$), Egal , and Util welfare
Although finding an allocation that maximizes the utilitarian welfare is straightforward (by allocating each item to the agent who has highest value), finding such an allocation within fair allocations is NP-hard \cite{conf/atal/BarmanG0KN19}. %journals/corr/abs-1908-00161 
One interesting research question here is to bound the utilitarian welfare loss by enforcing the allocations to be fair, i.e., the {\em price of fairness} \cite{journals/mst/BeiLMS21,conf/wine/BarmanB020}.
Besides utilitarian welfare, a large body of works studied the compatibility between fairness notions and the weaker efficiency notion of PO.
% One more attractive allocation is the one that is both fair and efficient. 
% recent focus in the field of fair allocation is to study whether certain fairness property is compatible with economic efficiencies, e.g., the Pareto optimality (PO).
For the case of goods, \citet{journals/teco/CaragiannisKMPS19} proved that the allocation that maximizes Nash welfare 
%(the product of all agents' utilities) 
is EF1 and PO.
Later, \citet{conf/sigecom/BarmanKV18} designed a pseudopolynomial time algorithm for computing EF1+PO allocations.
Truly polynomial time algorithms for the problem
%for computing such an allocation 
remain unknown.
\citet{conf/aaai/BarmanK19} designed a polynomial time algorithm for computing PROP1+PO allocations.
Regarding the stronger fairness notion of EFX, \citet{journals/tcs/AmanatidisBFHV21} proved that for bi-valued valuations, the allocation that maximizes Nash welfare is EFX+PO, and \citet{conf/sagt/GargM21} improved this result by giving a polynomial time algorithm. 
Further, \citet{conf/sagt/GargM21} proved that if the valuations have three different values, EFX+PO allocations may not exist.
In contrast, \citet{conf/aaai/HosseiniSVX21} proved that if the valuations are lexicographic, EFX+PO allocations exist and can be found in polynomial time.
% Regarding approximate MMS, since Pareto improvements preserve the MMS approximation, the compatibility of approximate MMS and PO is straightforward, but polynomial time algorithm i.
% However, it remains unknown whether polynomial time algorithms to compute these fair and efficient allocations, e.g., EF1 and PO, $0.5$-MMS and PO exist.

% Since Pareto improvements preserve the MMS approximation of an allocation, the compatibility of approximate MMS and PO is also known.
% However, it remains unknown whether polynomial time algorithms to compute these fair and efficient allocations, e.g., EF1 and PO, $0.5$-MMS and PO exist.

For chores, most of the problems are still open. 
The good news is that PROP1+PO allocations can be computed in polynomial time, even if the items are mixture of goods and chores \cite{journals/orl/AzizMS20}.
% It is known that PROP1 allocation always exists and can be found in polynomial time when the items are goods \cite{conf/sigecom/ConitzerF017,conf/aaai/BarmanK19}, chores \cite{journals/corr/abs-1907-01766} or mixture of goods and chores \cite{journals/orl/AzizMS20}.
% Moreover, all these results compute allocations that are both PROP1 and PO.
However, for EF1 and PROPX, their compatibility with PO are still unknown. 
% There are some recent progresses. 
% on the compatibility of EF1 and PO. 
\citet{journals/corr/abs-2110-11285} and \citet{journals/corr/abs-2110-09601} proved that for bi-valued instances,  EF1+PO allocations always exist and can be found efficiently, which are the only exceptions so far.
% Later \citet{journals/corr/PropxPO} showed that the stronger version of PROPX is not compatible with PO but the weaker version is under bi-valued and lexicographic valuations.

%even with only goods, a PROPX allocation may not exist \cite{moulin2018fair,journals/orl/AzizMS20}.
%When all items are chores, a PROPX allocation exists and can be found efficiently \cite{moulin2018fair,journals/corr/abs-2103-11849}. 
%However, the compatibility between PROPX and PO is still unknown.

% For the case of chores, it has recently been shown that for bi-valued instances there exist polynomial time algorithms for the computation of EF1 and PO allocations~\cite{journals/corr/abs-2110-09601}, and MMS and PO allocations~\cite{journals/corr/abs-2110-11285}.
% However, the existence and computation of EF1 and PO allocations for general instances remains unknown.
% It is also open whether polynomial time algorithms for the computation of approximate MMS and PO allocations exist.
% There are also works that consider compatibility of relaxations of proportionality and PO~\cite{journals/corr/abs-1907-01766,journals/corr/abs-2005-04864,journals/orl/AzizMS20}.

% \begin{itemize}
% 	\item Poly-time algorithm for WPROP1+PO for chores on fixed number of agents
	
% 	\item (Generalized) Leximin is PROP1+PO for $n=3,4$~\cite{journals/corr/abs-2005-04864}
	
% 	\item PROP1+PO for mixture of goods and chores~\cite{journals/orl/AzizMS20}
	
% 	\item FPTAS for CEEI for chores?~\cite{journals/corr/abs-2107-06649} 
% \end{itemize}

\subsection{Being Fair for Strategic Agents}

Fair allocation problems are often faced by {\em strategic} agents in real-life scenarios, where an agent may intentionally misreport her values for the items to manipulate the outcome of the algorithm and obtain a bundle of higher value. 
% The mechanism design perspective (in which the agents may misreport their preferences to manipulate the allocation) of the fair allocation problem is also a very interesting future direction.
The goal is to design {\em truthful} algorithms in which agents maximize their utilities by reporting true preferences.
For two agents, \citet{conf/sigecom/AmanatidisBCM17} gave a complete characterization of truthful algorithms,
using which we have the tight approximation bounds for solution concepts such as EF1 and MMS.
For an arbitrary number of agents, \citet{conf/ijcai/AmanatidisBM16} and \citet{conf/ijcai/00010W19} designed truthful approximation algorithms but the tight bounds are still unknown. 
The aforementioned works consider the social environment where monetary transfers are not allowed. % in the design of algorithms. 
With monetary transfers, polynomial time truthful mechanisms were designed in \cite{conf/atal/BarmanG0KN19} for single-parameter valuations, which maximize the social welfare and approximately satisfy fairness notions such as MMS and EF1.

Another game-theoretic research agenda is to investigate the agents' strategic behaviours in the algorithms that may not be truthful.  
For example, \citet{conf/wine/AmanatidisBFLLR21} proved that in the Round-Robin algorithm, the allocations induced by pure Nash equilibria are always EF1 (regarding the true values). 
\citet{conf/ecai/BouveretL14} and \citet{conf/aaai/AzizBLM17} studied the strategic setting in general sequential allocation algorithms.
It is interesting to study the agents' behaviours in other algorithms and with other fairness notions. 

% It is still unknown how to extend this result to arbitrary number of agents and the case of chores. 

% The best known approximations for MMS allocations by strategyproof algorithms are $\lfloor (m-n+2)/2 \rfloor$ for goods~\cite{conf/ijcai/AmanatidisBM16} and $O(\log\frac{m}{n})$ for chores~\cite{conf/ijcai/00010W19}.
% While the approximation for goods cannot be improved, the best possible approximation for chores remains largely open: the best known lower bound is $\Omega(1)$~\cite{conf/ijcai/00010W19}.
% It is also unknown whether the \emph{picking-exchange mechanism} for two agents~\cite{conf/sigecom/AmanatidisBCM17} can be extended to general number of agents to enhance our understanding of strategyproof allocation mechanisms (for both goods and chores).
% Finally, it would be interesting to study strategyproof mechanisms for approximating other fairness notions, e.g., EF1 and EFX~\cite{conf/atal/BarmanG0KN19,conf/wine/AmanatidisBFLLR21}.

% \bo{Is there any mechanism that can use money to achieve truthfulness and fairness? I am only aware of VCG mechanisms which maximizes social welfare}

{
\small
\bibliographystyle{named}
\bibliography{chores}

\begin{thebibliography}{}

\bibitem[\protect\citeauthoryear{Aleksandrov and
  Walsh}{2020}]{conf/aaai/AleksandrovW20}
Martin Aleksandrov and Toby Walsh.
\newblock Online fair division: {A} survey.
\newblock In {\em {AAAI}}, 2020.

\bibitem[\protect\citeauthoryear{Amanatidis \bgroup \em et al.\egroup
  }{2016}]{conf/ijcai/AmanatidisBM16}
Georgios Amanatidis, Georgios Birmpas, and Evangelos Markakis.
\newblock On truthful mechanisms for maximin share allocations.
\newblock In {\em {IJCAI}}, pages 31--37, 2016.

\bibitem[\protect\citeauthoryear{Amanatidis \bgroup \em et al.\egroup
  }{2017a}]{conf/sigecom/AmanatidisBCM17}
Georgios Amanatidis, Georgios Birmpas, George Christodoulou, and Evangelos
  Markakis.
\newblock Truthful allocation mechanisms without payments: Characterization and
  implications on fairness.
\newblock In {\em {EC}}, pages 545--562, 2017.

\bibitem[\protect\citeauthoryear{Amanatidis \bgroup \em et al.\egroup
  }{2017b}]{journals/talg/AmanatidisMNS17}
Georgios Amanatidis, Evangelos Markakis, Afshin Nikzad, and Amin Saberi.
\newblock Approximation algorithms for computing maximin share allocations.
\newblock {\em {ACM} Trans. Algorithms}, 13(4):52:1--52:28, 2017.

\bibitem[\protect\citeauthoryear{Amanatidis \bgroup \em et al.\egroup
  }{2018}]{conf/ijcai/AmanatidisBM18}
Georgios Amanatidis, Georgios Birmpas, and Vangelis Markakis.
\newblock Comparing approximate relaxations of envy-freeness.
\newblock In {\em {IJCAI}}, pages 42--48, 2018.

\bibitem[\protect\citeauthoryear{Amanatidis \bgroup \em et al.\egroup
  }{2020}]{journals/tcs/AmanatidisMN20}
Georgios Amanatidis, Evangelos Markakis, and Apostolos Ntokos.
\newblock Multiple birds with one stone: Beating 1/2 for {EFX} and {GMMS} via
  envy cycle elimination.
\newblock {\em Theor. Comput. Sci.}, 841:94--109, 2020.

\bibitem[\protect\citeauthoryear{Amanatidis \bgroup \em et al.\egroup
  }{2021a}]{journals/tcs/AmanatidisBFHV21}
Georgios Amanatidis, Georgios Birmpas, Aris Filos{-}Ratsikas, Alexandros
  Hollender, and Alexandros~A. Voudouris.
\newblock Maximum nash welfare and other stories about {EFX}.
\newblock {\em Theor. Comput. Sci.}, 863:69--85, 2021.

\bibitem[\protect\citeauthoryear{Amanatidis \bgroup \em et al.\egroup
  }{2021b}]{conf/wine/AmanatidisBFLLR21}
Georgios Amanatidis, Georgios Birmpas, Federico Fusco, Philip Lazos, Stefano
  Leonardi, and Rebecca Reiffenh{\"{a}}user.
\newblock Allocating indivisible goods to strategic agents: Pure nash
  equilibria and fairness.
\newblock In {\em {WINE}}, volume 13112, pages 149--166, 2021.

\bibitem[\protect\citeauthoryear{Aziz and
  Brandl}{2020}]{journals/corr/abs-2008-08991}
Haris Aziz and Florian Brandl.
\newblock The vigilant eating rule: {A} general approach for probabilistic
  economic design with constraints.
\newblock {\em CoRR}, abs/2008.08991, 2020.

\bibitem[\protect\citeauthoryear{Aziz and Mackenzie}{2016}]{conf/focs/AzizM16}
Haris Aziz and Simon Mackenzie.
\newblock A discrete and bounded envy-free cake cutting protocol for any number
  of agents.
\newblock In {\em {FOCS}}, pages 416--427, 2016.

\bibitem[\protect\citeauthoryear{Aziz \bgroup \em et al.\egroup
  }{2015}]{DBLP:journals/ai/AzizGMW15}
Haris Aziz, Serge Gaspers, Simon Mackenzie, and Toby Walsh.
\newblock Fair assignment of indivisible objects under ordinal preferences.
\newblock {\em Artif. Intell.}, 227:71--92, 2015.

\bibitem[\protect\citeauthoryear{Aziz \bgroup \em et al.\egroup
  }{2017a}]{conf/aaai/AzizBLM17}
Haris Aziz, Sylvain Bouveret, J{\'{e}}r{\^{o}}me Lang, and Simon Mackenzie.
\newblock Complexity of manipulating sequential allocation.
\newblock In {\em {AAAI}}, pages 328--334, 2017.

\bibitem[\protect\citeauthoryear{Aziz \bgroup \em et al.\egroup
  }{2017b}]{conf/aaai/AzizRSW17}
Haris Aziz, Gerhard Rauchecker, Guido Schryen, and Toby Walsh.
\newblock Algorithms for max-min share fair allocation of indivisible chores.
\newblock In {\em {AAAI}}, pages 335--341, 2017.

\bibitem[\protect\citeauthoryear{Aziz \bgroup \em et al.\egroup
  }{2019a}]{conf/ijcai/0001C019}
Haris Aziz, Hau Chan, and Bo~Li.
\newblock Weighted maxmin fair share allocation of indivisible chores.
\newblock In {\em {IJCAI}}, pages 46--52, 2019.

\bibitem[\protect\citeauthoryear{Aziz \bgroup \em et al.\egroup
  }{2019b}]{conf/ijcai/00010W19}
Haris Aziz, Bo~Li, and Xiaowei Wu.
\newblock Strategyproof and approximately maxmin fair share allocation of
  chores.
\newblock In {\em {IJCAI}}, pages 60--66, 2019.

\bibitem[\protect\citeauthoryear{Aziz \bgroup \em et al.\egroup
  }{2020a}]{journals/corr/abs-2012-13884}
Haris Aziz, Bo~Li, and Xiaowei Wu.
\newblock Approximate and strategyproof maximin share allocation of chores with
  ordinal preferences.
\newblock {\em CoRR}, abs/2012.13884, 2020.

\bibitem[\protect\citeauthoryear{Aziz \bgroup \em et al.\egroup
  }{2020b}]{journals/orl/AzizMS20}
Haris Aziz, Herv{\'{e}} Moulin, and Fedor Sandomirskiy.
\newblock A polynomial-time algorithm for computing a pareto optimal and almost
  proportional allocation.
\newblock {\em Oper. Res. Lett.}, 48(5):573--578, 2020.

\bibitem[\protect\citeauthoryear{Aziz \bgroup \em et al.\egroup
  }{2022}]{journals/aamas/AzizCIW22}
Haris Aziz, Ioannis Caragiannis, Ayumi Igarashi, and Toby Walsh.
\newblock Fair allocation of indivisible goods and chores.
\newblock {\em Auton. Agents Multi Agent Syst.}, 36(1):3, 2022.

\bibitem[\protect\citeauthoryear{Aziz}{2020a}]{conf/aaai/Aziz20}
Haris Aziz.
\newblock Developments in multi-agent fair allocation.
\newblock In {\em {AAAI}}, pages 13563--13568, 2020.

\bibitem[\protect\citeauthoryear{Aziz}{2020b}]{conf/wine/Aziz20}
Haris Aziz.
\newblock Simultaneously achieving ex-ante and ex-post fairness.
\newblock In {\em {WINE}}, volume 12495, pages 341--355, 2020.

\bibitem[\protect\citeauthoryear{Babaioff \bgroup \em et al.\egroup
  }{2021a}]{conf/sigecom/BabaioffEF21}
Moshe Babaioff, Tomer Ezra, and Uriel Feige.
\newblock Fair-share allocations for agents with arbitrary entitlements.
\newblock In {\em {EC}}, page 127, 2021.

\bibitem[\protect\citeauthoryear{Babaioff \bgroup \em et al.\egroup
  }{2021b}]{journals/mor/BabaioffNT21}
Moshe Babaioff, Noam Nisan, and Inbal Talgam{-}Cohen.
\newblock Competitive equilibrium with indivisible goods and generic budgets.
\newblock {\em Math. Oper. Res.}, 46(1):382--403, 2021.

\bibitem[\protect\citeauthoryear{Bansal and
  Sviridenko}{2006}]{conf/stoc/BansalS06}
Nikhil Bansal and Maxim Sviridenko.
\newblock The santa claus problem.
\newblock In {\em {STOC}}, pages 31--40, 2006.

\bibitem[\protect\citeauthoryear{Barman and
  Krishnamurthy}{2019}]{conf/aaai/BarmanK19}
Siddharth Barman and Sanath~Kumar Krishnamurthy.
\newblock On the proximity of markets with integral equilibria.
\newblock In {\em {AAAI}}, pages 1748--1755, 2019.

\bibitem[\protect\citeauthoryear{Barman and
  Krishnamurthy}{2020}]{journals/teco/BarmanK20}
Siddharth Barman and Sanath~Kumar Krishnamurthy.
\newblock Approximation algorithms for maximin fair division.
\newblock {\em {ACM} Trans. Economics and Comput.}, 8(1):5:1--5:28, 2020.

\bibitem[\protect\citeauthoryear{Barman \bgroup \em et al.\egroup
  }{2018a}]{conf/aaai/BarmanBMN18}
Siddharth Barman, Arpita Biswas, Sanath Kumar~Krishna Murthy, and Yadati
  Narahari.
\newblock Groupwise maximin fair allocation of indivisible goods.
\newblock In {\em {AAAI}}, 2018.

\bibitem[\protect\citeauthoryear{Barman \bgroup \em et al.\egroup
  }{2018b}]{conf/sigecom/BarmanKV18}
Siddharth Barman, Sanath~Kumar Krishnamurthy, and Rohit Vaish.
\newblock Finding fair and efficient allocations.
\newblock In {\em {EC}}, pages 557--574. {ACM}, 2018.

\bibitem[\protect\citeauthoryear{Barman \bgroup \em et al.\egroup
  }{2019}]{conf/atal/BarmanG0KN19}
Siddharth Barman, Ganesh Ghalme, Shweta Jain, Pooja Kulkarni, and Shivika
  Narang.
\newblock Fair division of indivisible goods among strategic agents.
\newblock In {\em {AAMAS}}, 2019.

\bibitem[\protect\citeauthoryear{Barman \bgroup \em et al.\egroup
  }{2020}]{conf/wine/BarmanB020}
Siddharth Barman, Umang Bhaskar, and Nisarg Shah.
\newblock Optimal bounds on the price of fairness for indivisible goods.
\newblock In {\em {WINE}}, pages 356--369, 2020.

\bibitem[\protect\citeauthoryear{Bei \bgroup \em et al.\egroup
  }{2021a}]{journals/ai/BeiLLLL21}
Xiaohui Bei, Zihao Li, Jinyan Liu, Shengxin Liu, and Xinhang Lu.
\newblock Fair division of mixed divisible and indivisible goods.
\newblock {\em Artif. Intell.}, 293:103436, 2021.

\bibitem[\protect\citeauthoryear{Bei \bgroup \em et al.\egroup
  }{2021b}]{journals/aamas/BeiLLW21}
Xiaohui Bei, Shengxin Liu, Xinhang Lu, and Hongao Wang.
\newblock Maximin fairness with mixed divisible and indivisible goods.
\newblock {\em Auton. Agents Multi Agent Syst.}, 35(2):34, 2021.

\bibitem[\protect\citeauthoryear{Bei \bgroup \em et al.\egroup
  }{2021c}]{journals/mst/BeiLMS21}
Xiaohui Bei, Xinhang Lu, Pasin Manurangsi, and Warut Suksompong.
\newblock The price of fairness for indivisible goods.
\newblock {\em Theory Comput. Syst.}, 65(7):1069--1093, 2021.

\bibitem[\protect\citeauthoryear{Berger \bgroup \em et al.\egroup
  }{2021}]{journals/corr/abs-2102-10654}
Ben Berger, Avi Cohen, Michal Feldman, and Amos Fiat.
\newblock (almost full) {EFX} exists for four agents (and beyond).
\newblock {\em CoRR}, abs/2102.10654, 2021.

\bibitem[\protect\citeauthoryear{Bhaskar \bgroup \em et al.\egroup
  }{2021}]{conf/approx/BhaskarSV21}
Umang Bhaskar, A.~R. Sricharan, and Rohit Vaish.
\newblock On approximate envy-freeness for indivisible chores and mixed
  resources.
\newblock In {\em {APPROX-RANDOM}}, 2021.

\bibitem[\protect\citeauthoryear{Bogomolnaia and
  Moulin}{2001}]{journals/jet/BogomolnaiaM01}
Anna Bogomolnaia and Herv{\'{e}} Moulin.
\newblock A new solution to the random assignment problem.
\newblock {\em J. Econ. Theory}, 100(2):295--328, 2001.

\bibitem[\protect\citeauthoryear{Bogomolnaia \bgroup \em et al.\egroup
  }{2017}]{BogomolnaiaMSY17}
Anna Bogomolnaia, Hervé Moulin, Fedor Sandomirskiy, and Elena Yanovskaya.
\newblock {Competitive Division of a Mixed Manna}.
\newblock {\em Econometrica}, 85(6):1847--1871, 2017.

\bibitem[\protect\citeauthoryear{Bogomolnaia \bgroup \em et al.\egroup
  }{2019}]{journals/scw/BogomolnaiaMSY19a}
Anna Bogomolnaia, Herv{\'{e}} Moulin, Fedor Sandomirskiy, and Elena Yanovskaya.
\newblock Dividing goods and bads under additive utilities.
\newblock {\em Soc Choice Welf}, 52:395–417, 2019.

\bibitem[\protect\citeauthoryear{Boodaghians \bgroup \em et al.\egroup
  }{2021}]{journals/corr/abs-2107-06649}
Shant Boodaghians, Bhaskar~Ray Chaudhury, and Ruta Mehta.
\newblock Polynomial time algorithms to find an approximate competitive
  equilibrium for chores.
\newblock {\em CoRR}, abs/2107.06649, 2021.

\bibitem[\protect\citeauthoryear{Bouveret and
  Lang}{2011}]{conf/ijcai/BouveretL11}
Sylvain Bouveret and J{\'{e}}r{\^{o}}me Lang.
\newblock A general elicitation-free protocol for allocating indivisible goods.
\newblock In {\em {IJCAI}}, pages 73--78, 2011.

\bibitem[\protect\citeauthoryear{Bouveret and
  Lang}{2014}]{conf/ecai/BouveretL14}
Sylvain Bouveret and J{\'{e}}r{\^{o}}me Lang.
\newblock Manipulating picking sequences.
\newblock In {\em {ECAI}}, pages 141--146, 2014.

\bibitem[\protect\citeauthoryear{Brams and Taylor}{1996}]{books/daglib/0017730}
Steven~J. Brams and Alan~D. Taylor.
\newblock {\em Fair division - from cake-cutting to dispute resolution}.
\newblock 1996.

\bibitem[\protect\citeauthoryear{Brams and Taylor}{2000}]{books/daglib/0017729}
Steven~J. Brams and Alan~D. Taylor.
\newblock {\em The win-win solution - guaranteeing fair shares to everybody}.
\newblock 2000.

\bibitem[\protect\citeauthoryear{Br{\^{a}}nzei and
  Sandomirskiy}{2019}]{journals/corr/abs-1907-01766}
Simina Br{\^{a}}nzei and Fedor Sandomirskiy.
\newblock Algorithms for competitive division of chores.
\newblock {\em CoRR}, abs/1907.01766, 2019.

\bibitem[\protect\citeauthoryear{Brustle \bgroup \em et al.\egroup
  }{2020}]{conf/sigecom/BrustleDNSV20}
Johannes Brustle, Jack Dippel, Vishnu~V. Narayan, Mashbat Suzuki, and Adrian
  Vetta.
\newblock One dollar each eliminates envy.
\newblock In {\em {EC}}, pages 23--39, 2020.

\bibitem[\protect\citeauthoryear{Budish}{2011}]{journals/bqgt/Budish10}
Eric Budish.
\newblock The combinatorial assignment problem: Approximate competitive
  equilibrium from equal incomes.
\newblock {\em Journal of Political Economy}, 119(6):1061--1103, 2011.

\bibitem[\protect\citeauthoryear{Caragiannis and
  Ioannidis}{2021}]{conf/wine/CaragiannisI21}
Ioannis Caragiannis and Stavros Ioannidis.
\newblock Computing envy-freeable allocations with limited subsidies.
\newblock In {\em {WINE}}, volume 13112, pages 522--539, 2021.

\bibitem[\protect\citeauthoryear{Caragiannis \bgroup \em et al.\egroup
  }{2019a}]{conf/ec/CaragiannisGH19}
Ioannis Caragiannis, Nick Gravin, and Xin Huang.
\newblock Envy-freeness up to any item with high nash welfare: The virtue of
  donating items.
\newblock In {\em {EC}}, pages 527--545, 2019.

\bibitem[\protect\citeauthoryear{Caragiannis \bgroup \em et al.\egroup
  }{2019b}]{journals/teco/CaragiannisKMPS19}
Ioannis Caragiannis, David Kurokawa, Herv{\'{e}} Moulin, Ariel~D. Procaccia,
  Nisarg Shah, and Junxing Wang.
\newblock The unreasonable fairness of maximum nash welfare.
\newblock {\em {ACM} Trans. Economics and Comput.}, 7(3):12:1--12:32, 2019.

\bibitem[\protect\citeauthoryear{Chakraborty \bgroup \em et al.\egroup
  }{2020}]{conf/atal/ChakrabortyISZ20}
Mithun Chakraborty, Ayumi Igarashi, Warut Suksompong, and Yair Zick.
\newblock Weighted envy-freeness in indivisible item allocation.
\newblock In {\em {AAMAS}}, pages 231--239, 2020.

\bibitem[\protect\citeauthoryear{Chakraborty \bgroup \em et al.\egroup
  }{2021}]{journals/corr/abs-2112-04166}
Mithun Chakraborty, Erel Segal{-}Halevi, and Warut Suksompong.
\newblock Weighted fairness notions for indivisible items revisited.
\newblock {\em CoRR}, abs/2112.04166, 2021.

\bibitem[\protect\citeauthoryear{Chaudhury \bgroup \em et al.\egroup
  }{2020}]{conf/sigecom/ChaudhuryGM20}
Bhaskar~Ray Chaudhury, Jugal Garg, and Kurt Mehlhorn.
\newblock {EFX} exists for three agents.
\newblock In {\em {EC}}, 2020.

\bibitem[\protect\citeauthoryear{Chaudhury \bgroup \em et al.\egroup
  }{2021a}]{conf/soda/ChaudhuryGMM21}
Bhaskar~Ray Chaudhury, Jugal Garg, Peter McGlaughlin, and Ruta Mehta.
\newblock Competitive allocation of a mixed manna.
\newblock In {\em {SODA}}, pages 1405--1424, 2021.

\bibitem[\protect\citeauthoryear{Chaudhury \bgroup \em et al.\egroup
  }{2021b}]{conf/sigecom/ChaudhuryGMMM21}
Bhaskar~Ray Chaudhury, Jugal Garg, Kurt Mehlhorn, Ruta Mehta, and Pranabendu
  Misra.
\newblock Improving {EFX} guarantees through rainbow cycle number.
\newblock In {\em {EC}}, 2021.

\bibitem[\protect\citeauthoryear{Chaudhury \bgroup \em et al.\egroup
  }{2021c}]{conf/aaai/ChaudhuryGM21}
Bhaskar~Ray Chaudhury, Jugal Garg, and Ruta Mehta.
\newblock Fair and efficient allocations under subadditive valuations.
\newblock In {\em {AAAI}}, pages 5269--5276, 2021.

\bibitem[\protect\citeauthoryear{Chaudhury \bgroup \em et al.\egroup
  }{2021d}]{journals/siamcomp/ChaudhuryKMS21}
Bhaskar~Ray Chaudhury, Telikepalli Kavitha, Kurt Mehlhorn, and Alkmini
  Sgouritsa.
\newblock A little charity guarantees almost envy-freeness.
\newblock {\em {SIAM} J. Comput.}, 2021.

\bibitem[\protect\citeauthoryear{Conitzer \bgroup \em et al.\egroup
  }{2017}]{conf/sigecom/ConitzerF017}
Vincent Conitzer, Rupert Freeman, and Nisarg Shah.
\newblock Fair public decision making.
\newblock In {\em {EC}}, 2017.

\bibitem[\protect\citeauthoryear{Conitzer \bgroup \em et al.\egroup
  }{2019}]{conf/aaai/ConitzerF0V19}
Vincent Conitzer, Rupert Freeman, Nisarg Shah, and Jennifer~Wortman Vaughan.
\newblock Group fairness for the allocation of indivisible goods.
\newblock In {\em {AAAI}}, pages 1853--1860, 2019.

\bibitem[\protect\citeauthoryear{Dehghani \bgroup \em et al.\egroup
  }{2018}]{conf/soda/DehghaniFHY18}
Sina Dehghani, Alireza Farhadi, Mohammad~Taghi Hajiaghayi, and Hadi Yami.
\newblock Envy-free chore division for an arbitrary number of agents.
\newblock In {\em {SODA}}, 2018.

\bibitem[\protect\citeauthoryear{Dubins and Spanier}{1961}]{dubins1961cut}
Lester~E Dubins and Edwin~H Spanier.
\newblock How to cut a cake fairly.
\newblock {\em The American Mathematical Monthly}, 68(1P1):1--17, 1961.

\bibitem[\protect\citeauthoryear{Ebadian \bgroup \em et al.\egroup
  }{2021}]{journals/corr/abs-2110-11285}
Soroush Ebadian, Dominik Peters, and Nisarg Shah.
\newblock How to fairly allocate easy and difficult chores.
\newblock {\em CoRR}, abs/2110.11285, 2021.

\bibitem[\protect\citeauthoryear{Edward~Su}{1999}]{edward1999rental}
Francis Edward~Su.
\newblock Rental harmony: Sperner's lemma in fair division.
\newblock {\em The American mathematical monthly}, 106(10):930--942, 1999.

\bibitem[\protect\citeauthoryear{Fain \bgroup \em et al.\egroup
  }{2018}]{conf/sigecom/FainM018}
Brandon Fain, Kamesh Munagala, and Nisarg Shah.
\newblock Fair allocation of indivisible public goods.
\newblock In {\em {EC}}, 2018.

\bibitem[\protect\citeauthoryear{Farhadi \bgroup \em et al.\egroup
  }{2019}]{journals/jair/FarhadiGHLPSSY19}
Alireza Farhadi, Mohammad Ghodsi, Mohammad~Taghi Hajiaghayi, S{\'{e}}bastien
  Lahaie, David~M. Pennock, Masoud Seddighin, Saeed Seddighin, and Hadi Yami.
\newblock Fair allocation of indivisible goods to asymmetric agents.
\newblock {\em J. Artif. Intell. Res.}, 64:1--20, 2019.

\bibitem[\protect\citeauthoryear{Feige \bgroup \em et al.\egroup
  }{2021}]{journals/corr/abs-2104-04977}
Uriel Feige, Ariel Sapir, and Laliv Tauber.
\newblock A tight negative example for {MMS} fair allocations.
\newblock In {\em {WINE}}, pages 355--372, 2021.

\bibitem[\protect\citeauthoryear{Foley}{1966}]{foley1966resource}
Duncan~Karl Foley.
\newblock {\em Resource allocation and the public sector}.
\newblock Yale University, 1966.

\bibitem[\protect\citeauthoryear{Freeman \bgroup \em et al.\egroup
  }{2020}]{conf/sigecom/Freeman0V20}
Rupert Freeman, Nisarg Shah, and Rohit Vaish.
\newblock Best of both worlds: Ex-ante and ex-post fairness in resource
  allocation.
\newblock In {\em {EC}}, pages 21--22, 2020.

\bibitem[\protect\citeauthoryear{Gafni \bgroup \em et al.\egroup
  }{2021}]{journals/corr/abs-2109-08671}
Yotam Gafni, Xin Huang, Ron Lavi, and Inbal Talgam{-}Cohen.
\newblock Unified fair allocation of goods and chores via copies.
\newblock {\em CoRR}, abs/2109.08671, 2021.

\bibitem[\protect\citeauthoryear{Garg and Murhekar}{2021}]{conf/sagt/GargM21}
Jugal Garg and Aniket Murhekar.
\newblock Computing fair and efficient allocations with few utility values.
\newblock In {\em {SAGT}}, volume 12885 of {\em Lecture Notes in Computer
  Science}, pages 345--359. Springer, 2021.

\bibitem[\protect\citeauthoryear{Garg and Taki}{2021}]{journals/ai/GargT21}
Jugal Garg and Setareh Taki.
\newblock An improved approximation algorithm for maximin shares.
\newblock {\em Artif. Intell.}, 300:103547, 2021.

\bibitem[\protect\citeauthoryear{Garg \bgroup \em et al.\egroup
  }{2019}]{conf/soda/GargMT19}
Jugal Garg, Peter McGlaughlin, and Setareh Taki.
\newblock Approximating maximin share allocations.
\newblock In {\em {SOSA}}, volume~69, pages 20:1--20:11, 2019.

\bibitem[\protect\citeauthoryear{Garg \bgroup \em et al.\egroup
  }{2021a}]{conf/sagt/GargHMS21}
Jugal Garg, Martin Hoefer, Peter McGlaughlin, and Marco Schmalhofer.
\newblock When dividing mixed manna is easier than dividing goods: Competitive
  equilibria with a constant number of chores.
\newblock In {\em {SAGT}}, volume 12885, pages 329--344, 2021.

\bibitem[\protect\citeauthoryear{Garg \bgroup \em et al.\egroup
  }{2021b}]{journals/corr/abs-2110-09601}
Jugal Garg, Aniket Murhekar, and John Qin.
\newblock Fair and efficient allocations of chores under bivalued preferences.
\newblock {\em CoRR}, abs/2110.09601, 2021.

\bibitem[\protect\citeauthoryear{Ghodsi \bgroup \em et al.\egroup
  }{2018}]{conf/sigecom/GhodsiHSSY18}
Mohammad Ghodsi, Mohammad~Taghi Hajiaghayi, Masoud Seddighin, Saeed Seddighin,
  and Hadi Yami.
\newblock Fair allocation of indivisible goods: Improvements and
  generalizations.
\newblock In {\em {EC}}, pages 539--556, 2018.

\bibitem[\protect\citeauthoryear{Halpern and Shah}{2019}]{conf/sagt/HalpernS19}
Daniel Halpern and Nisarg Shah.
\newblock Fair division with subsidy.
\newblock In {\em {SAGT}}, pages 374--389, 2019.

\bibitem[\protect\citeauthoryear{Halpern and Shah}{2021}]{conf/ijcai/0002021}
Daniel Halpern and Nisarg Shah.
\newblock Fair and efficient resource allocation with partial information.
\newblock In {\em {IJCAI}}, pages 224--230, 2021.

\bibitem[\protect\citeauthoryear{Hill}{1987}]{hill1987partitioning}
Theodore~P Hill.
\newblock Partitioning general probability measures.
\newblock {\em The Annals of Probability}, pages 804--813, 1987.

\bibitem[\protect\citeauthoryear{Hosseini \bgroup \em et al.\egroup
  }{2021a}]{DBLP:conf/aaai/HosseiniSVX21}
Hadi Hosseini, Sujoy Sikdar, Rohit Vaish, and Lirong Xia.
\newblock Fair and efficient allocations under lexicographic preferences.
\newblock In {\em {AAAI}}, pages 5472--5480, 2021.

\bibitem[\protect\citeauthoryear{Hosseini \bgroup \em et al.\egroup
  }{2021b}]{conf/aaai/HosseiniSVX21}
Hadi Hosseini, Sujoy Sikdar, Rohit Vaish, and Lirong Xia.
\newblock Fair and efficient allocations under lexicographic preferences.
\newblock In {\em {AAAI}}, pages 5472--5480, 2021.

\bibitem[\protect\citeauthoryear{Huang and Lu}{2021}]{conf/sigecom/HuangL21}
Xin Huang and Pinyan Lu.
\newblock An algorithmic framework for approximating maximin share allocation
  of chores.
\newblock In {\em {EC}}, pages 630--631, 2021.

\bibitem[\protect\citeauthoryear{Kulkarni \bgroup \em et al.\egroup
  }{2021}]{conf/sigecom/KulkarniMT21}
Rucha Kulkarni, Ruta Mehta, and Setareh Taki.
\newblock Indivisible mixed manna: On the computability of {MMS+PO}
  allocations.
\newblock In {\em {EC}}, pages 683--684, 2021.

\bibitem[\protect\citeauthoryear{Kurokawa \bgroup \em et al.\egroup
  }{2018}]{journals/jacm/KurokawaPW18}
David Kurokawa, Ariel~D. Procaccia, and Junxing Wang.
\newblock Fair enough: Guaranteeing approximate maximin shares.
\newblock {\em J. {ACM}}, 65(2):8:1--8:27, 2018.

\bibitem[\protect\citeauthoryear{Lang and Rothe}{2016}]{books/sp/16/LangR16}
J{\'{e}}r{\^{o}}me Lang and J{\"{o}}rg Rothe.
\newblock Fair division of indivisible goods.
\newblock In {\em Economics and Computation}, pages 493--550. 2016.

\bibitem[\protect\citeauthoryear{Lenstra \bgroup \em et al.\egroup
  }{1990}]{journals/mp/LenstraST90}
Jan~Karel Lenstra, David~B. Shmoys, and {\'{E}}va Tardos.
\newblock Approximation algorithms for scheduling unrelated parallel machines.
\newblock {\em Math. Program.}, 46:259--271, 1990.

\bibitem[\protect\citeauthoryear{Li \bgroup \em et al.\egroup
  }{2022}]{journals/corr/abs-2103-11849}
Bo~Li, Yingkai Li, and Xiaowei Wu.
\newblock Almost proportional allocations for indivisible chores.
\newblock {\em TheWebConf}, 2022.

\bibitem[\protect\citeauthoryear{Lipton \bgroup \em et al.\egroup
  }{2004}]{conf/sigecom/LiptonMMS04}
Richard~J. Lipton, Evangelos Markakis, Elchanan Mossel, and Amin Saberi.
\newblock On approximately fair allocations of indivisible goods.
\newblock In {\em {EC}}, pages 125--131, 2004.

\bibitem[\protect\citeauthoryear{Moulin}{1990}]{moulin1990uniform}
Herv{\'e} Moulin.
\newblock Uniform externalities: Two axioms for fair allocation.
\newblock {\em J. Public Econ.}, 43(3):305--326, 1990.

\bibitem[\protect\citeauthoryear{Moulin}{2003}]{books/daglib/0017734}
Herv{\'{e}} Moulin.
\newblock {\em Fair division and collective welfare}.
\newblock {MIT} Press, 2003.

\bibitem[\protect\citeauthoryear{Moulin}{2018}]{moulin2018fair}
Herv{\'e} Moulin.
\newblock Fair division in the age of internet.
\newblock {\em Annu. Rev. Econ.}, 2018.

\bibitem[\protect\citeauthoryear{Nisan}{2000}]{conf/sigecom/Nisan00}
Noam Nisan.
\newblock Bidding and allocation in combinatorial auctions.
\newblock In {\em {EC}}, pages 1--12, 2000.

\bibitem[\protect\citeauthoryear{Oh \bgroup \em et al.\egroup
  }{2021}]{journals/siamdm/OhPS21}
Hoon Oh, Ariel~D. Procaccia, and Warut Suksompong.
\newblock Fairly allocating many goods with few queries.
\newblock {\em {SIAM} J. Discret. Math.}, 35(2):788--813, 2021.

\bibitem[\protect\citeauthoryear{Plaut and
  Roughgarden}{2020}]{journals/siamdm/PlautR20}
Benjamin Plaut and Tim Roughgarden.
\newblock Almost envy-freeness with general valuations.
\newblock {\em {SIAM} J. Discret. Math.}, 34(2):1039--1068, 2020.

\bibitem[\protect\citeauthoryear{Robertson and
  Webb}{1998}]{books/daglib/0017738}
Jack~M. Robertson and William~A. Webb.
\newblock {\em Cake-cutting algorithms - be fair if you can}.
\newblock 1998.

\bibitem[\protect\citeauthoryear{Steinhaus}{1948}]{steinhaus1948problem}
Hugo Steinhaus.
\newblock The problem of fair division.
\newblock {\em Econometrica}, 16:101--104, 1948.

\bibitem[\protect\citeauthoryear{Suksompong}{2018}]{journals/mss/Suksompong18}
Warut Suksompong.
\newblock Approximate maximin shares for groups of agents.
\newblock {\em Math. Soc. Sci.}, 92:40--47, 2018.

\bibitem[\protect\citeauthoryear{Suksompong}{2021}]{suksompong2021constraints}
Warut Suksompong.
\newblock Constraints in fair division.
\newblock {\em ACM SIGecom Exchanges}, 19(2):46--61, 2021.

\bibitem[\protect\citeauthoryear{Sun \bgroup \em et al.\egroup
  }{2021}]{conf/atal/SunCD21}
Ankang Sun, Bo~Chen, and Xuan~Vinh Doan.
\newblock Connections between fairness criteria and efficiency for allocating
  indivisible chores.
\newblock In {\em {AAMAS}}, pages 1281--1289, 2021.

\bibitem[\protect\citeauthoryear{Tinbergen}{1930}]{tinbergen1930mathematiese}
Jan Tinbergen.
\newblock Mathematiese psychologie.
\newblock {\em Mens en Maatschappij}, 6(4):342--352, 1930.

\bibitem[\protect\citeauthoryear{Varian}{1973}]{varian1973equity}
Hal~R Varian.
\newblock Equity, envy, and efficiency.
\newblock 1973.

\bibitem[\protect\citeauthoryear{Walsh}{2020}]{conf/ijcai/Walsh20}
Toby Walsh.
\newblock Fair division: The computer scientist's perspective.
\newblock In {\em {IJCAI}}, pages 4966--4972, 2020.

\bibitem[\protect\citeauthoryear{Zhou and
  Wu}{2021}]{journals/corr/abs-2109-07313}
Shengwei Zhou and Xiaowei Wu.
\newblock Approximately {EFX} allocations for indivisible chores.
\newblock {\em CoRR}, abs/2109.07313, 2021.

\end{thebibliography}

}

\end{document}